\documentclass[prb,showpacs,twocolumn]{revtex4}
\usepackage{graphicx}
\begin{document}
\title{Band structure and optical properties of opal photonic crystals}
\author{E. Pavarini}
\author{L.C. Andreani}
\author{C. Soci}
\author{M. Galli}
\author{F. Marabelli}
\affiliation{Istituto Nazionale per la Fisica della Materia
         and Dipartimento di Fisica ``A. Volta'',
         Universit\`a di Pavia, Via Bassi 6, 27100 Pavia, Italy}

\author{D. Comoretto}
\affiliation{Istituto Nazionale per la Fisica della Materia
         and Dipartimento di Chimica e Chimica Industriale,
         Universit\`a di Genova, via Dodecaneso 31, 16146 Genova, Italy}
\date{\today}
\begin{abstract}
A theoretical approach for the interpretation of reflectance
spectra of opal photonic crystals with fcc structure
and (111) surface orientation is presented.
It is based on the calculation of photonic bands
and density of states corresponding to a specified
angle of incidence in air.
The results yield a clear distinction
between diffraction in the direction of light
propagation by (111) family planes (leading to the
formation of a stop band) and diffraction in other directions
by higher-order planes (corresponding to the excitation
of photonic modes in the crystal).
Reflectance measurements on artificial opals
made of self-assembled polystyrene spheres
are analyzed according to the theoretical scheme
and give evidence of diffraction by higher-order
crystalline planes in the photonic structure.
\end{abstract}
\maketitle

\section{Introduction}

Three-dimensional photonic crystals are intensively studied
with the goal of achieving control of light propagation
and, especially, a complete photonic band gap in all
directions.\cite{yablonovitch87,john87,soukoulis_nato,ieee}
Among the possible structures, fcc-based systems like
opals and inverse opals are of great interest
as they can be produced with bottom-up approaches
based on casting from colloidal solutions,
possibly followed by infiltration and template
removal.\cite{wijnhoven98,zakhidov98,blanco00,vlasov01}
The direct opal structure (e.g., polystyrene or silica
spheres in air) has weak effects on the total
photonic density of states (DOS) but does produce
a well-defined stop band in the direction of light propagation.
If the structure crystallizes in an fcc lattice with (111)
surface orientation, as is usually the case,
normal-incidence reflectance or transmittance
probe the photonic gap in the $\Gamma$L direction
of the fcc Brillouin zone. Instead, the inverse opal structure
(e.g., air spheres in TiO$_2$ or in silicon)
produces a strong effect on the photonic DOS
and may even result in a complete photonic band gap
if the dielectric constant and the filling fraction
of the infiltrated material are large enough.\cite{sozuer92,busch98}

The optical properties of fcc colloidal photonic crystals have
been the subject of numerous experimental
\cite{vos96,tarhan96,vlasov97,bogomolov97,miguez97,miguez99,thijssen99,imhof99,megens99,reynolds99,vlasov00_pre,vlasov00_apl,schriemer00,vandriel00,vos00,romanov01,koenderink02,lopez02,astratov02,galisteo-lopez03a,galisteo-lopez03b,miguez04,galisteo-lopez04}
and theoretical
\cite{sozuer92,yannopapas97,busch98,reynolds99,li00,yannopapas01,ochiai01,lopez-tejeira02,wang03}
studies. Many experimental investigations are concerned with
properties of the lowest-order gap, in particular the attenuation
studied by normal-incidence transmission\cite{vlasov97,miguez97}
and the evolution of the stop band in reflectance/transmittance
spectra at varying angles of
incidence\cite{bogomolov97,miguez99,thijssen99,reynolds99,vandriel00,vos00,galisteo-lopez03b}
and as a function of polarization.\cite{galisteo-lopez03a}
Detailed studies of the high-energy region at normal incidence
have recently appeared.\cite{galisteo-lopez04} The role of
disorder in Bragg scattering and gap formation has been addressed
with measurements on multiple-domain\cite{vlasov00_pre,astratov02}
and single-domain crystals.\cite{vlasov00_apl,lopez02} Bragg
diffraction has also been studied in a transmission geometry by
Kossel line analysis.\cite{tarhan96} The change in effective
refractive index close to the first-order gap and in connection
with flat bands at high energies has also been
studied.\cite{imhof99,miguez04} Laser dyes in opal photonic
crystals show effects related to modified photonic density of
states in emission spectra\cite{megens99,schriemer00,koenderink02}
and to interaction of photonic modes with excitons.\cite{eradat02}
On the theoretical side, the formation of a complete band gap in
the inverse opal structure\cite{sozuer92,busch98} and its
fragility against disorder\cite{li00} are now well understood. The
interplay between photonic bands and reflection/transmission
properties has been addressed both with calculations of optical
spectra\cite{yannopapas97,reynolds99,astratov02} and superprism
effects \cite{ochiai01} as well as with symmetry
analysis.\cite{lopez-tejeira02} More recent works focus on the
effects of stacking faults on the optical
properties.\cite{yannopapas01,wang03}

A question of particular interest is the role of diffraction from
\begin{figure}[!ht]
\begin{center}
\rotatebox{-90}{\includegraphics[width=0.3\textwidth]{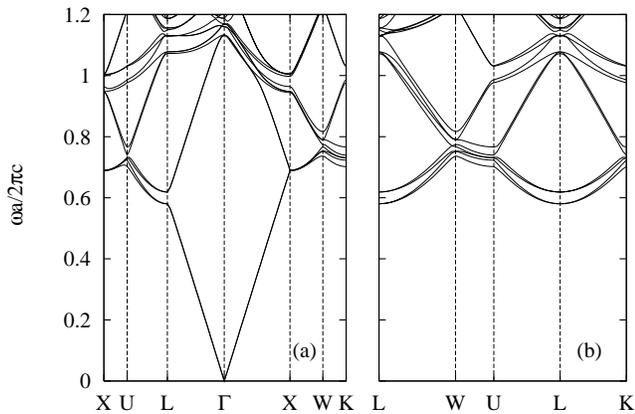}}
\vskip 1cm
\caption{Energy bands for an opal photonic crystal
         (fcc lattice of close-packed dielectric spheres
          with $\epsilon_1=2.53$ in air): (a) along symmetry directions
         in the whole Brillouin zone, (b) along symmetry directions
         on the hexagonal face of the Brillouin zone.}
\end{center}
\end{figure}
lattice planes other than the (111) family in connection with the
optical properties at energies higher than the first stop band
along the $\Gamma$L direction. The analysis of Kossel lines in
transmission has led to a determination of the photonic gap
positions for the two lowest band gaps\cite{tarhan96}: the results
for the band edges are reported as a function of incidence angle
along the LW direction of the fcc Brillouin zone. This
interpretation has been questioned in
Ref.~\onlinecite{yannopapas97} on the basis of calculations of
transmission spectra and projected photonic bands for given values
of the wavevector ${\mathbf{k}}_{\parallel}$ parallel to the
crystal surface. Angle-resolved reflectance measurements give
evidence of multiple diffraction peaks with small splittings at
large values of the angle of
incidence,\cite{vandriel00,vos00,romanov01,galisteo-lopez03b}
which have been interpreted as arising from multiple Bragg-wave
coupling\cite{vandriel00,vos00,galisteo-lopez03b} or photonic band
branching,\cite{romanov01} both concepts being related to an
anticrossing of photonic levels close to the U or W points of the
Brillouin zone. Related observations but a different
interpretation is given in Ref.\onlinecite{miguez04}, in which
additional peaks in reflectance (or dips in transmittance) at
near-normal incidence are attributed to a large effective
refractive index arising from flat bands. The main issue in the
interpretation of all these results is that spectral features in
optical measurements cannot be unambiguously associated with the
presence of a photonic gap and/or with diffraction. For
reflectance measurements, in particular, a spectral feature may
follow from the existence of a photonic gap (as is the case for
the lowest stop band at near-normal incidence) but also from the
excitation of a photonic mode in the crystal: establishing a clear
distinction between the two effects is the main purpose of the
present paper.

In this work we present a theoretical approach to the optical
properties of 3D photonic crystals which yields further insight in
the phenomenon of diffraction from different families of lattice
planes in relation to the presence of {\em photonic gaps} or {\em
photonic bands}. We calculate the photonic bands and density of
states corresponding to a fixed external angle of incidence: the
DOS is decreased when a photonic gap along the propagation
direction occurs, while it increases and shows resonant structures
when diffraction in directions other than that of specular
reflection takes place. Calculations of reflection and diffraction
spectra show that peaks in the DOS correspond to the onset of
diffraction, as well as to additional reflectance structures at
large values of the angle of incidence. Reflectance measurements
on opal photonic crystals are presented and interpreted with the
above concepts. In particular, weak features at large angles of
incidence give experimental evidence for the excitation of
photonic modes in the crystal, i.e., for diffraction from planes
other than those of the (111) family.

The results of this work are related to the surface
\begin{figure}[]
\vspace*{1cm}
\begin{center}
\rotatebox{0}{\includegraphics[width=0.4\textwidth]{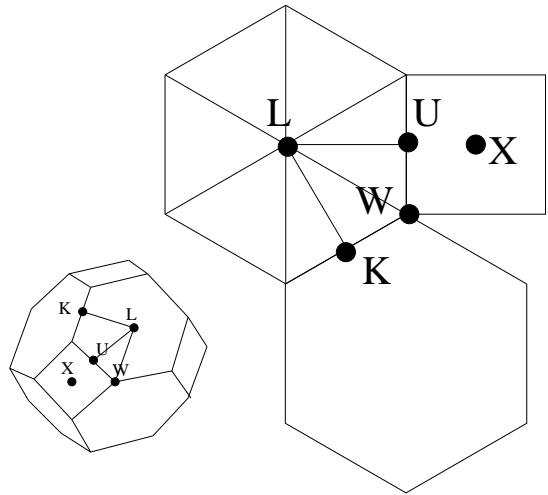}}
\vskip 0.5cm

\caption{Symmetry points in the Brillouin zone for a fcc lattice.}
\end{center}
\end{figure}

reflectance technique for measuring the photonic band dispersion.
This approach was first applied to photonic crystal waveguides,
where it led to a determination of the dispersion of
photonic modes above the light line
via the observation of resonant structure in reflectance
and of their dispersion with incidence
angle.\cite{fujita98,astratov99,pacradouni00,galli02_epjb}
The same technique of variable-angle reflectance
from the crystal surface can be applied
to the case of macroporous silicon,
which has a 2D photonic structure and can be considered
as homogeneous in the direction of the pores,
and in this case it yields the dispersion of
photonic bands in the 2D plane.\cite{galli02_prb}
The present work shows that, in the case of photonic crystals
with refractive index modulation in 3D,
Bragg diffraction off the direction of light propagation
also produces spectral structures in reflectance,
whose angular evolution is related to the photonic dispersion
in a plane parallel to the crystal surface.
Preliminary results have been recently presented.\cite{comoretto04_spie}

The rest of this paper is organized as follows.
In Sec.II we present the theoretical approach,
namely the calculation of photonic DOS
and photonic bands corresponding to a specified
angle of incidence. In Sec.III we describe
reflectance experiments on opal photonic crystals
and compare the theoretical predictions with the optical spectra.
In Section IV  the findings of this work are summarized.

\section{Photonic bands and density of states}
\begin{figure*}[!ht]
\vspace*{-5cm}
\begin{center}
\rotatebox{270}{\includegraphics[width=0.8\textwidth]{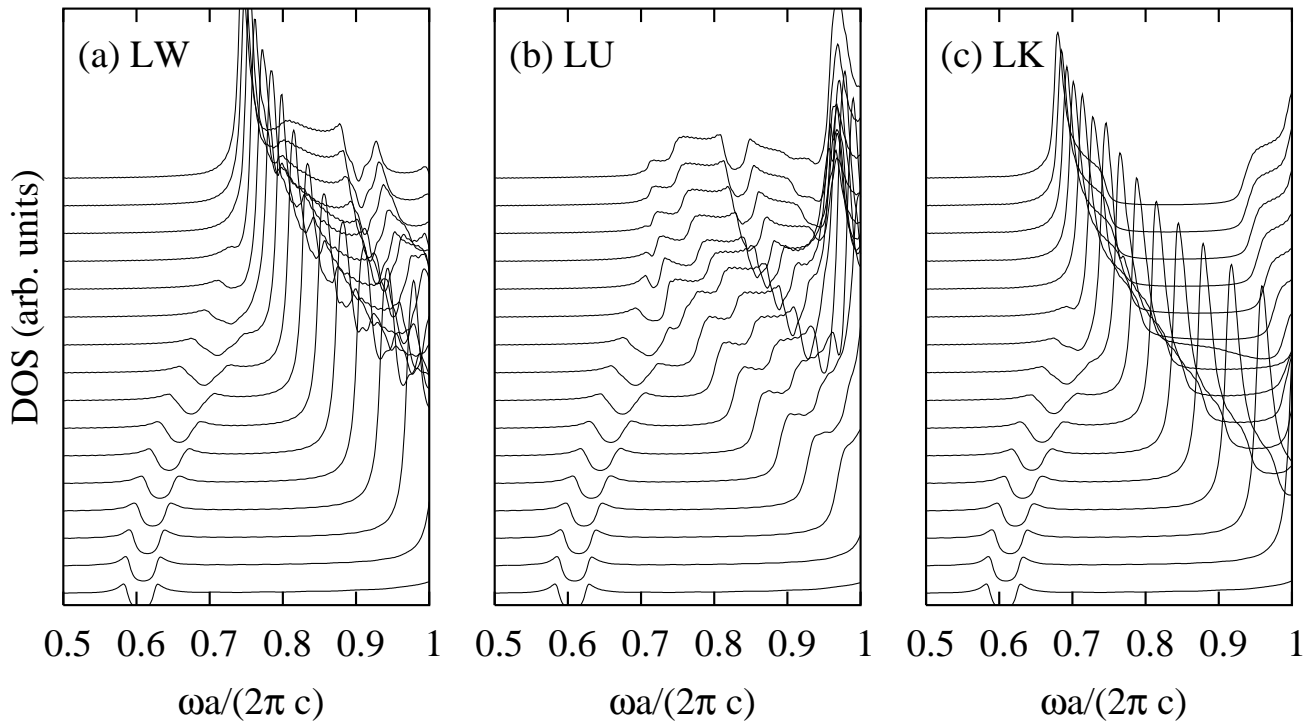}}
\caption
{Reduced density of states for several values of the incidence angle
 from
$\theta=5^\circ$ (bottom curve) to $\theta=80^\circ$ (top curve) in
steps of $5^\circ$.
The three figures correspond to different values
of the azimuthal angle $\phi$:
(a) $\phi=0^\circ + 2m\pi/6$ ($m \in \mathcal{Z}$), or LW direction;
(b) $\phi=30^\circ + 2m\pi/3$, or LU direction;
(c) $\phi=-30^\circ + 2m\pi/3$, or LK direction.
 }
\end{center}
\end{figure*}

We calculate the photonic bands
starting from the second-order wave equation for the magnetic field
\begin{equation}
\nabla\times\left[\frac{1}{\varepsilon(\mbox{\bf r})}
\nabla\times\mbox{\bf H}\right]
=\frac{\omega^2}{c^2}\mbox{\bf H}.
\end{equation}
We adopt the plane-wave expansion technique to transform Eq.~(1)
into an eigenvalue problem, and calculate the inverse dielectric constant
by using the procedure of Ho, Chan and Soukoulis.\cite{ho90}
In Fig.~1 we show the photonic bands for an opal
with dielectric constants
$\epsilon_1=2.53$ (spheres)\cite{ma03,note_dispersion}
and $\epsilon_2=1$ (voids), i.e.,  polystyrene spheres in air.
The fcc Brillouin zone with high symmetry points is shown in Fig.~2.
The bands are shown in dimensionless units $\omega a/(2\pi c)$,
where $a$ is the lattice constant,
along the symmetry lines of the full fcc Brillouin zone (Fig.~1a)
as well as along symmetry lines on the (111) hexagonal face (Fig.~1b):
the latter will be useful for comparing with density of states
and reflectance spectra at a finite angle of incidence.

The photonic bands in Fig.~1 do not exhibit any
complete gap in the full Brillouin zone (BZ).
However, partial gaps do appear along some symmetry
lines of the BZ; among these, the lowest energy gap
opens along $\Gamma$L around $\omega a/(2\pi c)\sim 0.6$.
Partial gaps behave as stop bands in reflectance spectra
when they occur in the direction of propagation.
Photonic modes along the propagation direction
also contribute to the optical properties and
may give rise to additional features in reflectance,
as discussed in the rest of this Section.

In order to distinguish reflectance maxima due to stop bands
from features which can instead be associated to photonic modes,
we need first of all to calculate the photonic band structure and the
density of states (DOS) along the direction of propagation,
i.e., for a specified value of the angle of incidence in air.
In the following we will refer to these bands and DOS
as {\em reduced bands} and {\em reduced DOS}, respectively.
The surface of a fcc colloidal crystal is known
to be a (111) crystallographic plane,
therefore angle-resolved reflectivity experiments are usually performed
as a function of the angle of incidence with respect to a (111) surface.
We adopt the same convention in our calculations
and define a reference frame as follows:  $\hat{z}$ is the
[1,1,1] or $\Gamma$L direction, $\hat{x}$ is the [1,0,-1]
(a LW) direction, and ${y}$ is the [1,-2,1] (a LK) direction.
We can therefore express the momentum of an incident photon
as follows:
$
  {\bf k}^i=\frac{\omega}{c}
  (\sin\theta\cos\phi,\sin\theta\sin\phi,\cos\theta),
$
where $\hbar\omega$ is the energy of the incident photon.
With our choice of axes, $\theta=0^\circ$ corresponds to normal
incidence ($\Gamma$L direction)
and the azimuthal angle $\phi$ is measured from the LW direction.
We stress that $\theta$ is the angle of incidence in air
(which differs from the angle of propagation
inside the opal,\cite{ochiai01}
the latter being dependent on the effective refractive index
and on the presence of additional propagating modes when
diffraction inside the material occurs).

The reduced bands and state densities
are obtained from the full band structure (Fig.~1)
by using conservations laws:
(i) parallel momentum conservation modulo a reciprocal
lattice vector, and
(ii) energy conservation.
For the photonic bands, these two conditions may be written as
\begin{eqnarray}
 {\bf k}_{\parallel} +{\bf G}_\parallel &=&\frac{\omega}{c}
 (\sin\theta\cos\phi,\sin\theta\sin\phi,0), \label{eq1} \\
 \epsilon_n({\bf k}_{\parallel},{\bf k}_z) &= &\hbar \omega,
 \label{eq2}
\end{eqnarray}
where ${\bf k}_\parallel\equiv{\bf k}^i_\parallel$
is the component of {\bf k} parallel to the (111) surface,
        ${\bf G_\parallel}$ is the parallel component of
any reciprocal lattice vector ${\bf G}$,
${\bf k}_z$ is the component of {\bf k} parallel to $\Gamma$L,
and $\epsilon_n({{\bf{k}_\parallel,\bf{k}}_z})$ is the
$n$-th energy band of a photon with momentum {\bf k}.
We point out that $\bf{k}_\parallel$ is fixed by the experimental
configuration. On the contrary ${\bf k}_z$ and the reduced bands
are obtained from the solutions of the system given by
Eqs.~(\ref{eq1}),(\ref{eq2}), which are found by a selfconsistent
procedure.
The reduced DOS is obtained from the reduced bands as follows:
\begin{equation}\label{dos}
N(\hbar\omega)
=
\sum_{n,{\bf{k}}_z\ge0}
\delta (\epsilon_n({{\bf k}_\parallel,{\bf k}_z}) -\hbar\omega),
\end{equation}
where {\bf k}$_\parallel$ is defined by Eq.~(\ref{eq1}).
The above $N(\hbar\omega)$ is a one-dimensional density of states,
because it is summed only over ${\bf k}_z$.\cite{path}
The DOS obtained in this way
(convoluted with a Lorentzian of width comparable to a typical
experimental resolution $\Delta\omega=10^{-2}\cdot2\pi c/a$
and averaged over photon polarizations)
is shown in Fig.\thinspace 3  for several
values of $\theta$ and some values of $\phi$
(directions LW, LU and LK).\cite{note_lklu}

Let us analyze first the results for $\theta=0^\circ$.
In this case the solution of Eqs.~(\ref{eq1}) and (\ref{eq2})
can be found very simply because {\bf k}$_\parallel=0$,
and it yields the photonic bands along the $\Gamma$L direction.
Thus we can directly compare the $\theta=0^\circ$ DOS in Fig.\thinspace 3
with the bands along  $\Gamma$L in Fig.~1a.
The lowest energy gap shown in Fig.~1 around $\omega a /(2\pi c) =0.6$
appears clearly in the reduced DOS
at the same values of the energy.
It arises from diffraction in the backwards direction
by planes of the (111) family.
The photonic bands at energies above the gap in Fig.~1a
correspond to a photon momentum $\bf{k}$ outside the first Brillouin zone;
the bands are folded into the first BZ by subtracting from {\bf k}
a reciprocal lattice vector $(2\pi/a)(1,1,1)$.
These folded bands give rise to a finite DOS
above the band gap in Fig.~3.

The reduced DOS reflects, of course, the energy dispersion
of the photonic bands along the propagation direction.
The latter is almost linear at low frequencies, and
thus gives rise to a nearly constant one-dimensional DOS.
For the case shown in Fig.~1a, the bands along $\Gamma$L
just above the gap at $\omega a/(2\pi c)=0.6$
exhibit also an almost linear dispersion,
thus the corresponding DOS in Fig.~3 remains flat above the gap.
However, in general, the folded bands may exhibit a non-linear dispersion
and can also show critical points (maxima, minima, or saddle points).
In three dimensions, a critical point results in a divergence
of the slope of the DOS (van Hove singularity).
In the present one-dimensional case
the DOS itself diverges at a van Hove singularity.
The very flat bands at energy around $\omega a /(2\pi c) =1.1$
(the fifth and sixth bands of the photonic structure)
show one of such critical points at L;
as a consequence, the corresponding reduced DOS at normal
incidence exhibits a Van Hove singularity around $\omega a /(2\pi c) =1.1$.
This critical point is out of the scale of Fig.~3
(it is shown in Fig.~4 below);
however, as we discuss in next paragraph,
it moves to lower energies when moving from normal incidence.

The complex, nonlinear dispersion of the photonic bands
results from the mixing of plane waves with different
reciprocal lattice vectors (in directions other than the (111) one)
in the photonic Bloch mode,
thus it implies the occurrence of diffraction in directions
other than that of beam propagation.
Notice that this phenomenon can occur
at any angle of incidence, even for $\theta=0^\circ$,
when the conservation of parallel crystal momentum
expressed by Eq (2)
can be satisfied with ${\bf G}_{\parallel}=0$.
For energies below $\omega a/(2\pi c)=1.5$,
the photonic bands are determined by reciprocal lattice vectors of
lower modulus, the relevant ones being (000),
the eight equivalent \{111\}
and the six equivalent \{200\} ones.
Thus the directions of diffraction
out of the propagation direction (forward or backward)
are determined by planes of equivalent \{111\} families
that are not parallel to the (111) crystal surface,
as well as by \{200\} families. At near-normal incidence,
diffraction directions just above the diffraction cutoff
are determined by ($\bar{1}$11), (1$\bar{1}$1),
(11$\bar{1}$) family planes: on the other hand,
\{200\} families become important at large angles
of incidence, as discussed below.
We conclude that the presence of a van Hove singularity
in the reduced DOS marks the onset of diffraction
from lattice planes other than those of the (111) family.

Let us analyze now the case of finite incidence angles.
\begin{figure*}[!ht]
\begin{center}
\rotatebox{0}{\includegraphics[width=0.9\textwidth]{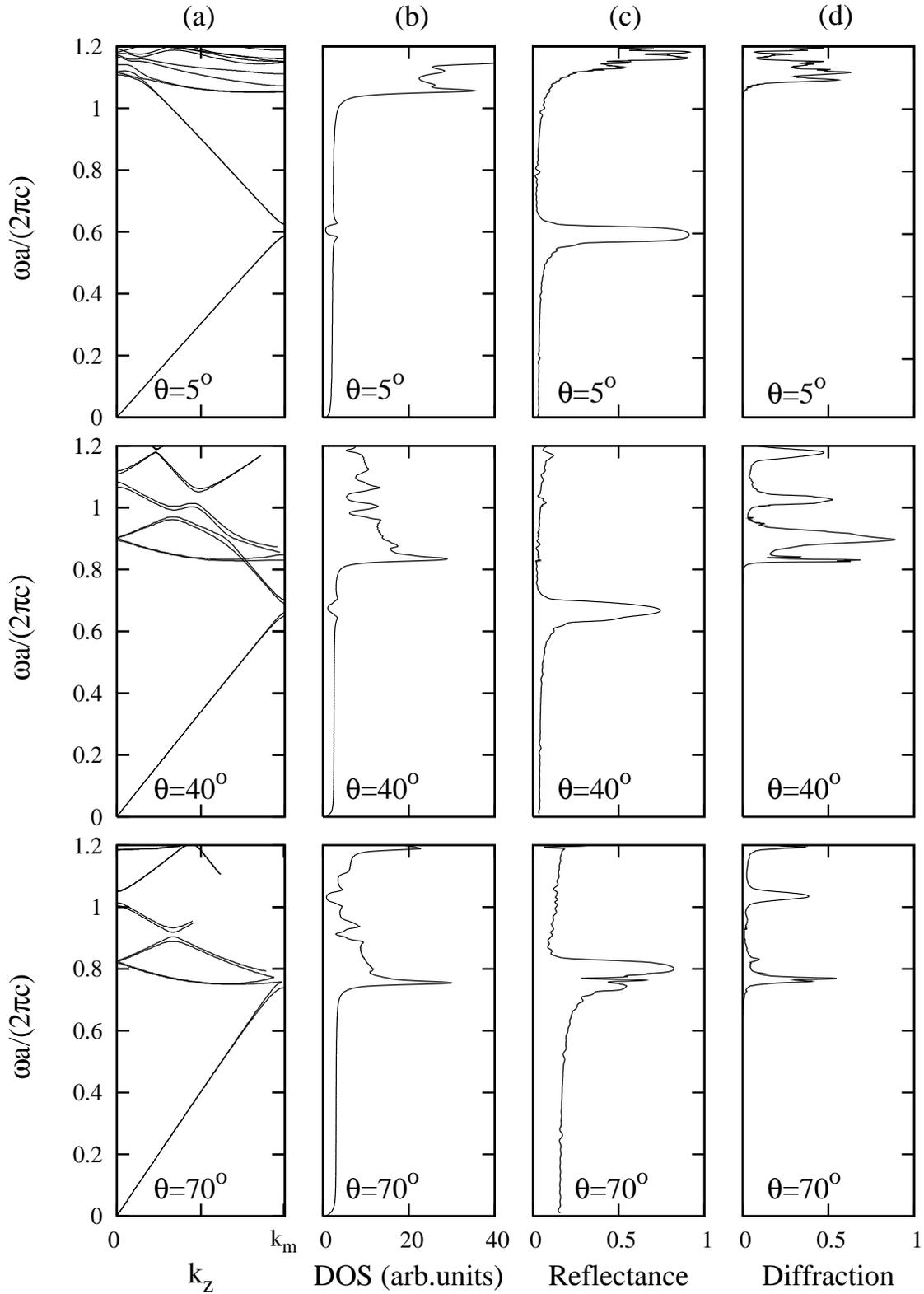}}
\vspace*{-2cm}
\caption{(a) Reduced bands, (b) reduced DOS, (c) reflectance and
(d) diffraction spectra for $\phi=0^\circ$ (LW direction)
and three values $\theta=5^\circ,\, 40^\circ\, 70^\circ$
of the angle of incidence in air.
The bands in (a) are plotted in the interval $(0,k_m)$,
where $k_m\equiv\sqrt{3}\pi/a$ is the $\Gamma$-L distance.
In the limit $\theta=0$ the point $k_m$ is identical to L.
The bands are displayed only for $k_z$ inside the first Brillouin zone.
 }
\end{center}
\end{figure*}
Figure~3 shows how the stop band depend strongly on the incidence angle,
by moving to higher energies with increasing $\theta$.
On the other hand, it depends only weakly on the azimuthal angle $\phi$.
Figure~3 shows also the peaks associated to the Van Hove singularity.
The latter is at $\omega a /(2\pi c) =1.1$ at normal incidence,
but it moves at lower energies as $\theta$ increases.
This is shown clearly in Figs.~3a and 3c,
in which this singularity appears as a strong peak in the DOS:
this peak merges with the gap around
$\theta=70^{\circ}$ and $60^{\circ}$, respectively.
Notice that since both the gap and the van Hove singularities
occur on the hexagonal face of the fcc Brillouin zone,
their angular dependence is related to the wavevector
dependence of the bands shown in Fig.~1b.
In particular, along the LW orientation,
the merging of the van Hove singularity
with the gap occurs close to the W point (see Fig.~1b),
where multiple Bragg scattering from \{111\} and
\{200\} family planes occurs.
We point out that the position of the Van Hove singularity
depends strongly on $\phi$,
as we may see by comparing Figs.~3a,c with Fig.~3b.
In particular, for $\phi=30^\circ$ (LU direction)
the Van Hove singularity remains almost outside the energy window.

Photonic modes give rise to several other features
in the reduced DOS. All these features are associated with folded bands
and exhibit a similar behavior as a function of $\theta$.
This can be seen clearly in Fig.~3b.
In this case the van Hove singularity remains very high in energy
for any value of $\theta$. The steps seen in the DOS are instead
associated with bands which have an almost linear dispersion;
a step indicates that a new band is becoming accessible.
Furthermore, Fig.~3b shows also a second photonic gap at higher energies,
which is present but much weaker
for $\phi=0^\circ$ (LW direction, Fig.~3a),
and is totally absent for $\phi=-30^\circ$ (LK direction, Fig.~3c).

In order to better clarify the $\theta$-dependence of
the stop bands and of the van Hove singularities,
in Fig.~4a,b we show the reduced bands and DOS in more detail.
The bands are shown as a function of ${\bf k}_z$
and for three values of $\theta$.
We focus on the LW direction ($\phi=0^\circ$),
but examination of the results
for other orientations leads to similar conclusions.
We notice that, since ${\bf k}_z$ is defined
only inside the first BZ, its maximum length
becomes shorter than $\sqrt{3}\pi/a$
for large $\theta$ and large frequencies (Fig.~4a for $\theta=70^\circ$),
due to the truncated shape of the BZ in Fig.~2:
this occurs when ${\bf k}_\parallel +{\bf G}_\parallel$
is outside the hexagonal face containing the L point.

For near-normal incidence, the bands along the direction of propagation
($\theta=5^\circ$) are simply the photonic bands along $\Gamma$L,
with the gap around $\omega a /(2\pi c)\sim 0.6$
and the flat bands around $\omega a /(2\pi c)\sim 1.1$.
The panels in Fig.~4a show that, when $\theta$ increases, the gap moves
to higher energy, while the flat bands and the associated
van Hove singularity move downwards to lower energy.
Eventually the gap and the flat bands
overlap for $\theta\sim 70^\circ$ and $\omega a/(2\pi c)\simeq0.75$,
corresponding to the W point  of the fcc Brillouin zone (see Fig.~1b).
These features are precisely reflected in the DOS,
as we may see in the panels of Fig.~4b.
In Fig.~4c,d we show reflectance and diffraction spectra
for the same values of the angle of incidence,
calculated by the scattering-matrix method.\cite{whittaker99,note_cylinder}
Diffraction (D) spectra are obtained as $D=1-R-T$ in terms
of reflectance (R) and transmittance (T).
The stop band appears clearly in reflectance spectra.
The onset of diffraction is
around $\omega a/(2\pi c)=1.1$ for $\theta=5^\circ$,
$\omega a/(2\pi c)=0.82$ for $\theta=40^\circ$
and $\omega a/(2\pi c)=0.75$ for $\theta=70^\circ$.
The first peak in diffraction spectra correlates very clearly
with the peak in the reduced DOS (Fig.~4b)
arising in turn from the flat bands and the Van Hove singularity (Fig.~4a).

We notice that other diffraction peaks and van Hove singularities
are present in Fig.4b,d for frequencies above the diffraction cutoff.
We do not pursue an analysis of these additional features, since
they have no well-resolved counterpart in either theoretical reflectance
spectra (Fig.4c) or in the experimental ones (see next Section),
and also because the approximation made in the theoretical
calculation becomes less accurate on increasing the energy.\cite{note_cylinder}
We emphasize that diffraction features become evident in
reflectance spectra at high values of the angle of incidence,
as clearly seen for $\theta=70^{\circ}$ in Fig.4c and
discussed below in connection with the experimental data.

Thus the analysis of the reduced bands and DOS corresponding to a
specified value of the angle of incidence leads to a clear
distinction between diffraction effects (high DOS regions) from
gaps or pseudogaps (zero or low DOS). In addition, critical points
may give rise to van Hove singularities which strongly enhance the
DOS in the associated spectral region. The presence of a van Hove
singularity marks the excitation of a photonic mode in the crystal
and corresponds to the occurrence of diffraction in directions
other than that of propagation. Van Hove singularities appear
clearly in diffraction spectra and become evident also in
reflection spectra at large values of the angle of incidence,
i.e. close to the W point,
where Bragg scattering is determined by planes of both \{111\} and
\{200\} families. These conclusions are important for interpreting
reflection experiments on polystyrene opals, as discussed in the
next Section.

\section{Optical spectra}

\subsection{Experimental}
Bulk polystyrene opals have been grown by monodisperse sphere
suspensions in water (Duke Scientific,
diameter 222, 260, 300, 340, 426 nm,
refractive index 1.59).\cite{ma03,note_dispersion}
Small vessels with area of about one squared centimeter
are obtained by sealing on a glass window polystyrene or silicone tubes.
These vessels are filled with the sphere suspension and placed in a
wet environment to reduce the evaporation rate.
The opals have a meniscus shape and a glossy surface
showing bright colored reflections.

Scanning Electron Microscopy (SEM) of a 222 nm opal surface
has been performed with a Leo Stereoscan
440 (LEO Electron Microscopy Ltd) electronic microscope.
The surface is very flat on a large scale (tens of microns)
and shows a regular arrangement of the spheres in a triangular lattice
as expected for their three-dimensional packing
in a fcc lattice when observed along the (111) direction.
Figure 5 shows a SEM image of 18$\times$12 $(\mu m)^2$ opal surface.
This image can be considered as representative
of the most disordered part of the opal surface investigated
by the optical measurements.
This morphology is very different
from that of opal films we previously investigated,
which were characterized by ordered domains of spheres
of a few square microns areas
with both triangular and square lattices,
tilted with respect to each other and separated by disordered
zones or deep dislocations.\cite{dav1,dav2}
For the present bulk opal samples, no square lattices are observed
and the surface appears as a large triangular lattice of spheres
with only a few defects.
\begin{figure}[!ht]
\begin{center}
\rotatebox{0}{\includegraphics[width=0.25\textwidth]{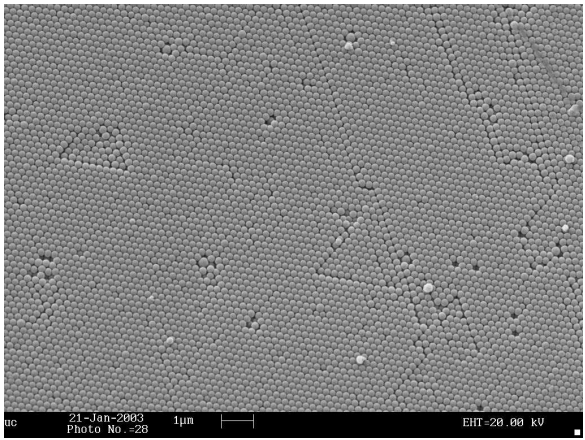}}
\vskip 1cm

\caption { SEM Image of the opal surface (18$\times$22 $\mu m^2$).}
\end{center}
\end{figure}

Unpolarized variable-angle specular reflectance has been measured
in the 0.4-4 eV spectral range by means of a
Fourier-Transform spectrophotometer
Bruker IFS66 with 8 cm$^{-1}$ spectral resolution.
The light of a Xe arc-lamp was collimated and then
focused to a spot of 100 $\mu m$ diameter on the sample surface.
The sample was placed on a home-made $\theta - 2 \theta$
goniometer, that allows the incidence angle to be
varied between 5 and 80 degrees, with an angular resolution of 1$^\circ$
determined by the angular divergence of the incident light cone.
Sample visualization and precise allocation of the light spot
on the surface was achieved by an optical microscope mounted on
the goniometer.
An Ag mirror was used as a reference. The overall error bar of
the experimental set-up on the absolute value of the
reflectance is within 2$\%$.

Particular care has been taken in the alignment of the sample.
Since the top quality surface of the investigated sample
is located close to one of the sharp border, we used the latter
as a macroscopic reference to identify the crystallographic axes
of the surface. As a matter of fact, atomic force microscope
images recorded at different magnifications showed that the LW
crystallographic direction has a well defined offset angle
with respect to the sharp border.
Therefore the sample was mounted on the microreflectometer
by aligning the incidence plane with the sharp border and
than rotating it by the offset angle.

\subsection{Results and discussion}

We apply our theoretical method to understand the
\begin{figure}[!ht]
\vspace*{-2.7cm}
\begin{center}
\rotatebox{0}{\includegraphics[width=0.6\textwidth]{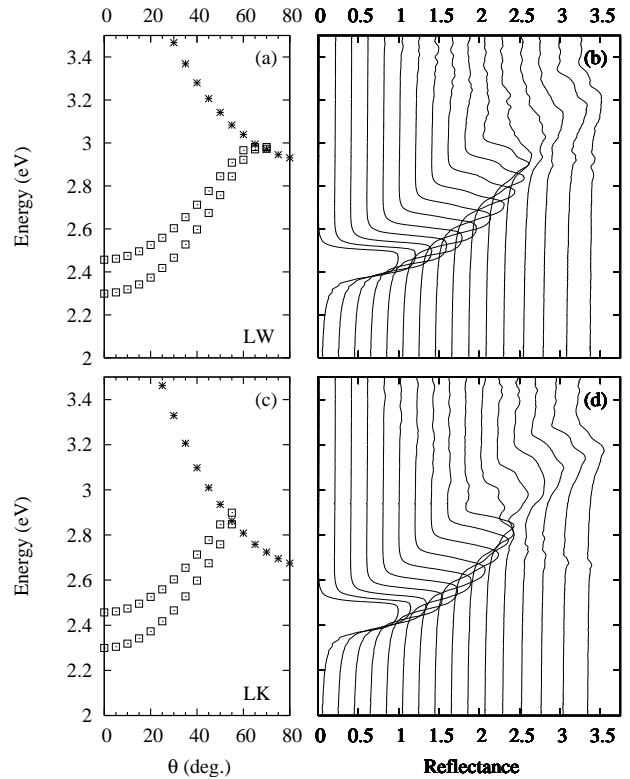}}
\caption {
  (ac):  Theoretical positions of the gap edges (squares) and of the
         lowest energy peak (stars) for the LW (a) and the
         LK (c) direction.
        The results are for a close-packed fcc structure
    of polystyrene spheres ($\epsilon=2.53$) with a lattice constant
    $a=314$~nm.
  (bd):  Experimental reflectance data for
    several values of $\theta=j\Delta\theta$, with
    $\Delta\theta=5^\circ$ for the LW (b) and the LK (d) direction;
    curves from left to right correspond to $j=1,\dots,16$
    and are shifted by the value $0.2(j-1)$.
}
\end{center}
\end{figure}

reflectance spectra for the sample shown in Fig. 5.
We therefore calculate the reduced bands for a close packed fcc crystal
with sphere diameter of 222 nm (fcc lattice constant $a=314$~nm).
We then obtain the reduced DOS and calculate the energy position
of the low energy gap and its dependence on $\theta$,
as well as the two most important high energy features,
i.e., two peaks that are associated with photonic modes.
The energy position of the gap
is obtained directly from the reduced bands (see Fig.~4),
while the position of the peaks is obtained from the DOS
shown in Fig.~3.
The results are shown in Fig.~6a and 6c  for the
directions LW and LK, respectively.
The angular dependence of these energy positions
correlates with the dispersion of the photonic bands
along the hexagonal surface of the Brillouin zone (Fig.~1b),
which gives the energies of Van Hove singularities.
Notice that the Van Hove singularity for the LK direction
has a lower energy (at a given value of the incidence angle)
as compared to the LW direction.
The Van Hove singularity crosses the band gap
at $\theta=70^\circ$ for LW and $\theta=60^\circ$ for LK.
The energies of these structures may be directly compared with the
features observed in the reflectance spectra (Fig.~6b and Fig.~6d).

Before starting a comparison of the theoretical
and experimental results, we give a description of the latter.
Figure 6b and 6d show the variable-angle micro reflectance spectra of the opal
for $\theta$ ranging from $5^\circ$ to $80^\circ$ in steps of 5$^\circ$
and along LW and LK, respectively.
These spectra are offset horizontally for a more clear view.
At near-normal incidence a single, intense
(reflectance almost identical to that of the Ag reference mirror),
asymmetrical band associated with the stop band
is observed at 2.47 eV with a FWHM of about 0.15 eV.

When the incidence angle is increased
\begin{figure}[!ht]
\vspace*{-.8cm}
\begin{center}
\rotatebox{270}{\includegraphics[width=0.7\textwidth]{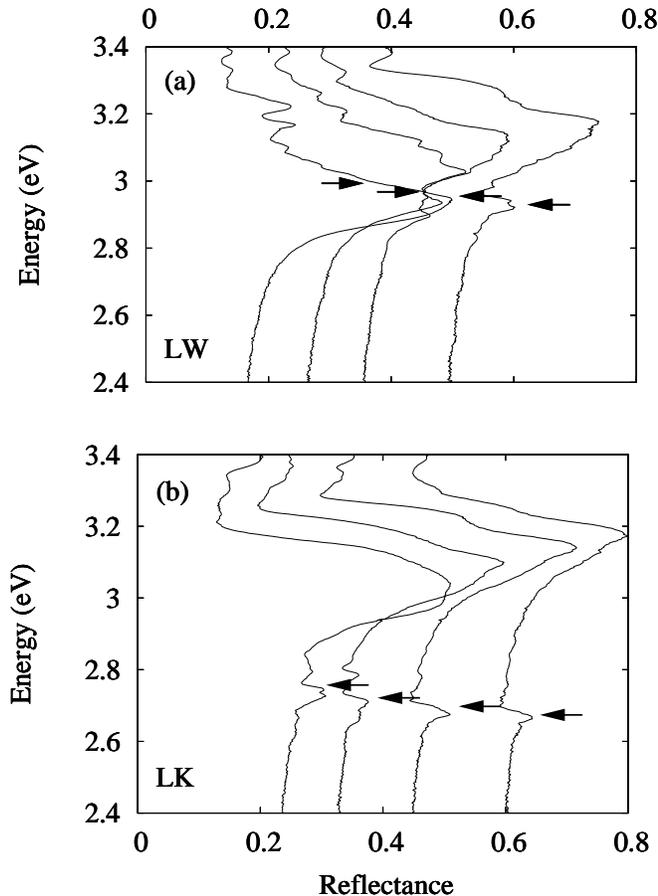}}

\caption { Enlarged view of experimental reflectance spectra
           for large incidence angles: $\theta=
           65^\circ, 70^\circ, 75^\circ, 80^\circ$
           (from left to right).
           Arrows indicate the calculated spectral positions
           of the lower energy Van Hove singularity
           (Fig.~6a and 6c).
           Directions are: LW (a) and LK (b).
}
\end{center}
\end{figure}
from normal incidence up to $45^\circ$,
the reflectance maximum shifts towards higher energies
slightly reducing its intensity and modifying its shape.
This behavior is accounted for by the theoretical
band edge positions summarized by the squares in Fig.~6a (LW)
and Fig.~6c (LK).
The agreement between calculated and measured gap position
as well as its dependence on $\theta$ is very satisfactory.

A more complex situation occurs for higher values of the incidence angle.
\begin{figure}[!ht]
\begin{center}
\rotatebox{0}{\includegraphics[width=0.5\textwidth]{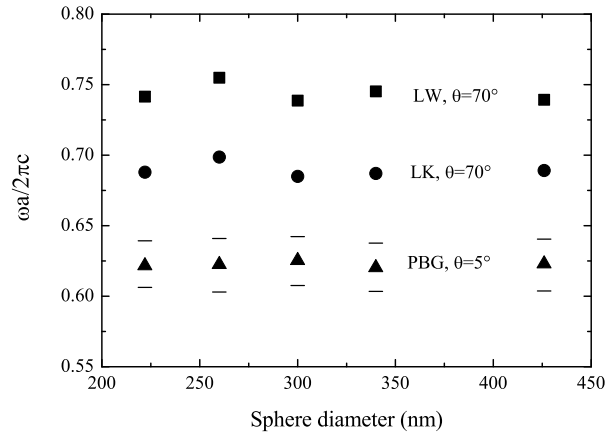}}

\caption {Scaling of the spectral features, in dimensionless units,
as a function of sphere diameter:
stop band central position (triangles)
and stop band edges (bars) at near-normal incidence;
Van Hove singularities at $\theta=70^{\circ}$
for LK (circles) and LW (squares) orientations.
          }
\end{center}
\end{figure}

We notice that for $\theta \ge 50^\circ$
the shape of the spectrum changes in a substantial way
and additional features occur.
The details of these features depend on the observed direction.
To give a better idea of the complicated structure of the spectra
recorded for large $\theta$ values and
in order to find evidence of the photonic modes,
we enlarge the energy scale
and show the relevant experimental curves in Fig.~7a (LW)
and Fig.~7b (LK).
A structured band is observed to regularly shift its maximum in the
range from 3 to 3.15 eV retaining the dispersion of the stop band.
It may roughly be assigned to the interplay of the stop band and
of the excitation of the photonic modes observed in the DOS
(Fig.~6a and Fig.~6c), which takes place for large $\theta$.
However we also notice the appearence of weak features
in the low-energy side of these bands.
In particular, for the LW direction the weak feature
is at about 2.9 eV, while for LK it is at about 2.7 eV.
When $\theta$ increases, these structures shift towards lower energies,
i.e., they behave like the calculated van Hove singularity.
Spectral features showing similar dispersions have
been previously observed in the transmission spectra of
polystyrene\cite{dav1}  and silica\cite{dav2} opal films.
The low energy calculated peak displayed in Fig.~6a and Fig.~6c
for high incidence angles is also reported as an arrow in Fig.~7a
and Fig.~7b to show a direct comparison with the experimental
spectra for each value of $\theta$.
We find a very good agreement between the calculated peaks
and the weak structures observed in the experimental spectra
for both LW and LK directions.
Moreover, the possibility of observing
a signature of diffraction effects in reflectance spectra
has been proved by the calculated spectra shown in Fig.4c.
We therefore assign the features having dispersion opposite
to the stop band to van Hove singularities in the reduced DOS,
i.e., to diffraction from family planes other than those of the (111) one.
For large values of the incidence angle along the LW orientation,
multiple Bragg scattering close to the W point takes place
and diffraction inside the opal is determined by \{111\}
as well as by \{200\} family planes.

In order to further support the assignment of the spectral features
observed in R spectra with van Hove singularities, we checked their
scaling with the lattice constant (related to the sphere diameter).
Figure 8 shows the experimental spectral position for the stop band
maximum and its width at near normal incidence, as well as van Hove
singularities measured along the LW and LK directions at $\theta=70^{\circ}$
for  opals with different sphere diameters. All these energies, expressed
in dimensionless units, are clearly independent of the lattice
constant thus unambiguously showing that the spectral structures
are due to intrinsic properties of the photonic crystal band structure.

It should be remarked that although the presence of flat bands is
essential in giving rise to van Hove singularities of the DOS and to
diffraction features, the interpretation of the present experiments
goes beyond that of Ref.\onlinecite{miguez04} in which related
observations are attributed only to a higher effective refractive
index arising from flat bands and diffraction is not considered.
Indeed, when diffraction starts to occur, the optical properties of
a photonic crystal cannot be described by a single effective index
as several modes propagate in the crystal at the same frequency. The
present interpretation does not contradict the multiple Bragg-wave
coupling or band branching mechanisms previously
introduced\cite{vandriel00,vos00,romanov01,galisteo-lopez03b} and it
confirms the identification of multiple diffraction as arising from
\{111\} or \{200\} family planes for the different orientations.
Extending the previous works, our observations indicate that
diffraction-related features in reflectance spectra become
increasingly visible for higher values of the angle of incidence:
i.e., the onset of diffraction (or, equivalently, the excitation of
several photonic modes at the same frequency) is not restricted to
the vicinity of the W or K/U points of the Brillouin zone and should
be viewed as a more general characteristic of the optical response
of the investigated photonic crystals.

We notice that weak spectral features with a negative dispersion
as a function of the incidence angle may also be recognized in the range
$30^\circ-60^\circ$ above the band gap peak (Fig. 6b and Fig. 6d).
We do not pursue a quantitative analysis of these features, whose
identification is not straightforward.
In fact photonic modes occuring at relatively high energy are more sensitive
to disorder, which may broaden and mix fine spectral features,
and also to dispersion of the refractive index.
Further work is required in order to establish conclusively
the presence of Van Hove singularities at energies above the band gap.

\section{Conclusions}

In this paper a detailed study of the photonic band structure
for artificial opals with close packed fcc structure is reported.
The reduced photonic bands and density of states
corresponding to a given value of the incidence angle in air
are calculated in order to provide a guide for the interpretation
of optical spectra.
Two main features with opposite dispersions have been identified.
A minimum of the reduced DOS corresponds to the stop band
in the propagation direction, while maxima and van Hove singularities
in the reduced DOS are associated with the excitation
of photonic modes in the crystal due to diffraction
by higher order planes in the photonic lattice.
These modes depend on the investigated
orientations of the sample with respect to the normal direction,
reflecting the microscopic symmetries of the photonic lattice.
By comparing with calculated reflection and diffraction spectra,
it has been shown that van Hove singularities
do indeed correspond to diffraction peaks
and become also visible in reflectance spectra
at large values of the incidence angle.

The theoretical data are compared with microreflectance
spectra recorded at different incidence angles
on bulk artificial opals.
There is a very good agreement between theoretical and
experimental data for what concerns the dispersion of the stop band.
The predicted dispersion of the excited photonic modes is
validated by features observed in reflectance spectra for large
incidence angles and different sample alignments along the
symmetry directions LW and LK,
thereby providing evidence for diffraction by higher-order planes
(namely \{111\} and \{200\} families),
or excitation of photonic modes out of the direction of light propagation.
Finally, all the spectral features here discussed and measured in
opals grown with different sphere diameter scale with the lattice
constant, thus confirming their assignment to intrinsic features
of the photonic band structure.
The role of light polarization on the optical properties
of opaline photonic crystals is currently under investigation.

\section{Acknowledgments}
The authors are indebted with Mario Agio for
providing the scattering-matrix code.
They are also grateful to D. Coquillat, J. Galisteo-L\'opez
and D. Wiersma  for helpful conversations.
D.C. thanks Prof. G. Dellepiane for her support.
This work was partially supported by the Ministry of Education,
University and Research (MIUR) through Cofin program.


\begin{thebibliography}{10}



\bibitem{yablonovitch87}
E. Yablonovitch, Phys. Rev. Lett. {\bf 58}, 2059 (1987).

\bibitem{john87}
S. John, Phys. Rev. Lett. {\bf 58}, 2486 (1987).

\bibitem{soukoulis_nato} {\em Photonic Crystals
and Light Localization in the 21st Century},
edited by C.M. Soukoulis, NATO Science Series C, vol. {\bf 563}
(Kluwer, Dordrecht, 2001).

\bibitem{ieee} For recent reviews, see e.g.\ papers
in IEEE J.\ Quantum Electron.\ $\mathbf{38}$,
Feature Section on Photonic Crystal Structures
and Applications, edited by T.F. Krauss and T. Baba,
pp. 724-963 (2002).

\bibitem{wijnhoven98} J.E.G.J. Wijnhoven and W.L. Vos,
Science {\bf 281}, 802 (1998).

\bibitem{zakhidov98} A.A. Zakhidov, R.H. Baughman, Z. Iqbal,
C. Cui, I. Khayrullin, S.O. Dantas, J.I. Marti,
and V.G. Ralchenko, Science {\bf 282}, 897 (1998).

\bibitem{blanco00} A. Blanco, E. Chomski, S. Grabtchak,
M. Ibisate, S. John, S.W. Leonard, C. L\'opez, F. Meseguer,
H. Miguez, J.P. Mondia, G.A. Ozin, O. Toader, and H.M. van Driel,
Nature {\bf 405}, 437 (2000).

\bibitem{vlasov01} Yu.A. Vlasov, X.-Z. Bo, J.C. Sturm, and D.J. Norris,
Nature {\bf 414}, 289 (2001).

\bibitem{sozuer92} H.S.\ S\"oz\"uer, J.W.\ Haus,
and N.\ Inguva, Phys.\ Rev.\ B {\bf 45} 13962 (1992).

\bibitem{busch98}
K.\ Busch, and S.\ John, Phys.\ Rev.\ E {\bf 58} 3896 (1998).



\bibitem{vos96}
W.L. Vos, R. Sprik, A. van Bladeren, A. Imhof,
A. Lagendijk, and G.H. Wegdam, Phys. Rev. B {\bf 53}, 16231 (1996).

\bibitem{tarhan96}
I.I. Tarhan and G.H. Watson,
Phys. Rev. Lett. {\bf 76}, 315 (1996).

\bibitem{vlasov97}
Yu.A. Vlasov, V.N. Astratov, O.Z. Karimov, A.A. Kaplyanskii,
V.N. Bogomolov, and A.V. Prokofiev,
Phys. Rev. B {\bf 55}, R13357 (1997).

\bibitem{bogomolov97}
V.N. Bogomolov, S.V. Gaponenko, I.N. Germanenko,
A.M. Kapitonov, E.P. Petrov, N.V. Gaponenko, A.V. Prokofiev,
A.N. Ponyanina, N.I. Silvanovich, and S.M. Samoilovich,
Phys. Rev. E {\bf 55}, 7619 (1997).

\bibitem{miguez97}
H. M\'{\i}guez, C. L\'opez, F. Meseguer, A. Blanco, L. V\'azquez,
R. Mayoral, M. Oca\~na, V. Forn\'es, and A. Mifsud,
Appl. Phys. Lett. {\bf 71}, 1148 (1997).

\bibitem{miguez99}
H. M\'{\i}guez, A. Blanco, F. Meseguer, C. L\'opez,
H.M. Yates, M.E. Pemble, V. Forn\'es, and A. Mifsud,
Phys. Rev. B {\bf 59}, 1563 (1999).

\bibitem{thijssen99} M. Thijssen, R. Sprik, J.E.G.J. Wijnhoven,
M. Megens, T. Narayanan, A. Lagendijk, and W.L. Vos,
Phys. Rev. Lett. {\bf 83}, 2730 (1999).

\bibitem{imhof99} A. Imhof, W.L. Vos, and A. Lagendijk,
Phys. Rev. Lett. {\bf 83}, 2942 (1999).

\bibitem{megens99}
M.M. Megens, J.E.G.J. Wijnhoven, A. Lagendijk, and W.L. Vos,
Phys. Rev. A {\bf 59}, 4727 (1999).

\bibitem{reynolds99}
A. Reynolds, F. L\'opez-Tejeira, D. Cassagne,
F.J. Garc\'{\i}a-Vidal, C. Jouanin, and J. S\'anchez-Dehesa,
Phys. Rev. B {\bf 60}, 11422 (1999).


\bibitem{vlasov00_pre}
Yu. A. Vlasov, V.N. Astratov, A.V. Baryshev, A.A. Kaplyanskii,
O.Z. Karimov, and M.F. Limonov,
Phys. Rev. E {\bf 61}, 5784 (2000).

\bibitem{vlasov00_apl}
Yu.A. Vlasov, M. Deutsch, and D.J. Norris,
Appl. Phys. Lett. {\bf 76}, 1627 (2000).

\bibitem{schriemer00}
H.P. Schriemer, H.M. van Driel, A.F. Koenderink, and W.L. Vos,
Phys. Rev. A {\bf 63}, 011801 (2000).

\bibitem{vandriel00}
H.M. van Driel and W.L. Vos, Phys. Rev. B {\bf 62}, 9872 (2000).

\bibitem{vos00}
W.L. Vos and H.M. van Driel, Phys. Lett. A {\bf 272}, 101 (2000).

\bibitem{romanov01}
S.G. Romanov, T. Maka, C.M. Sotomayor Torres, M. M\"uller,
R. Zentel, D. Cassagne, J. Manzanares-Martinez, and C. Jouanin,
Phys. Rev. E {\bf 63}, 056603 (2001).

\bibitem{koenderink02}
A.F. Koenderink, L. Bechger, H.P. Schriemer, A. Lagendijk,
and W.L. Vos, Phys. Rev. Lett. {\bf 88}, 143903 (2002).

\bibitem{lopez02}
J.F.G. L\`opez and W.L. Vos, Phys. Rev. E {\bf 66}, 036616 (2002).

\bibitem{astratov02} V.N. Astratov, A.M. Adawi, S. Fricker,
M.S. Skolnick, D.M. Whittaker, and P.N. Pusey,
Phys. Rev. B {\bf 66}, 165215 (2002).

\bibitem{galisteo-lopez03a} J.F. Galisteo-L\'opez, F. L\'{o}pez-Tejeira,
S. Rubio, C. L\'opez, and J. S\'anchez-Dehesa, Appl. Phys. Lett.
{\bf 82}, 4068 (2003).

\bibitem{miguez04} H. M\'{\i}guez, V. Kitaev, and G.A. Ozin,
Appl. Phys. Lett. {\bf 84}, 1239 (2004).

\bibitem{galisteo-lopez03b} J.F. Galisteo-L\'opez,
E. Palacios-Lid\'{o}n, E. Castillo-Mart\'{\i}nez, and C. L\'opez,
Phys. Rev. B {\bf 68}, 115109 (2003).

\bibitem{galisteo-lopez04} J.F. Galisteo-L\'opez and C. L\'opez,
Phys. Rev. B {\bf 70}, 035108 (2004).

\bibitem{yannopapas97}
V. Yannopapas, N. Stefanou, and A. Modinos,
J. Phys: Cond. Matt. {\bf 9}, 10261 (1997).

\bibitem{li00} Z.-Y. Li, and Z.-Q. Zhang,
Phys. Rev. B {\bf 62}, 1516 (2000).

\bibitem{yannopapas01}
V. Yannopapas, N. Stefanou, and A. Modinos,
Phys. Rev. Lett. {\bf 86}, 4811 (2001).

\bibitem{ochiai01}
T. Ochiai, K. Sakoda, and J. S\'anchez-Dehesa,
Phys. Rev. B {\bf 64}, 245113 (2001).

\bibitem{lopez-tejeira02}
F. L\'opez-Tejeira, T. Ochiai, K. Sakoda,
and J. S\'anchez-Dehesa, Phys. Rev. B {\bf 65}, 195110 (2002).

\bibitem{wang03}
Z.L. Wang, C.T. Chan, W.Y. Zhang, Z. Chen, N.B. Ming,
and P. Sheng, Phys. Rev. E {\bf 67}, 016612 (2003).

\bibitem{eradat02}
N. Eradat, A. Y. Sivachenko, M. E. Raikh, Z. V. Vardeny,
A. A. Zakhidov and R. H. Baughman,
Appl. Phys. Lett. {\bf 80},
3491 (2002).


\bibitem{fujita98}
T.~Fujita, Y.~Sato, T.~Kuitani, and T.~Ishihara,
Phys.~Rev.~B {\bf 57}, 12428 (1998).

\bibitem{astratov99}
V.N. Astratov, D.M. Whittaker, I.S. Culshaw, R.M. Stevenson,
M.S. Skolnick, T.F. Krauss, and R.M. De La Rue,
Phys.~Rev.~B {\bf 60}, R16255 (1999).

\bibitem{pacradouni00}
V. Pacradouni, W.J. Mandeville, A.R. Cowan, P. Paddon,
J.F. Young, and S.R. Johnson, Phys. Rev. B {\bf 62}, 4204 (2000).

\bibitem{galli02_epjb}
M. Galli, M. Agio, L.C. Andreani, L. Atzeni, D. Bajoni, G. Guizzetti, L.
Businaro, E. Di Fabrizio, F. Romanato, and A. Passaseo,
Eur. Phys. J. B {\bf 27}, 79 (2002).

\bibitem{galli02_prb}
M. Galli, M. Agio, L.C. Andreani, M. Belotti, G. Guizzetti,
F. Marabelli, M. Patrini, P. Bettotti, L. Dal Negro,
Z. Gaburro, L. Pavesi, A. Lui, and P. Bellutti,
Phys. Rev. B {\bf 65}, 113111 (2002).

\bibitem{comoretto04_spie}
D. Comoretto, E. Pavarini, M. Galli, C. Soci,
F. Marabelli, and L.C. Andreani,
SPIE Proc. {\bf 5511}, 135 (2004).

\bibitem{ho90} K.M.\ Ho, C.T.\ Chan, and C.M.\ Soukoulis,
Phys.\ Rev.\ Lett.\ {\bf 65}, (1990) 3152.

\bibitem{ma03} X. Ma, J.Q. Lu, R.S. Brock, K.M. Jacobs,
P. Yang, and X.-H. Hu, Phys. Med. Biol. {\bf 48}, 4165 (2003).

\bibitem{note_dispersion} According to Ref.~\onlinecite{ma03},
the refractive index $n=1.59$ of polystyrene spheres
refers to a wavelength $\lambda=500$ nm.
Its dispersion has been measured
to be no more than $\pm1.5\%$ around this value
in the wavelength range 400-1600 nm. The effect
of the index dispersion is negligibly small in
the frequency window relevant for spectral features
observed around 2.5-3 eV and discussed in this work.


\bibitem{path} For a given pair of values ($\theta,\phi$),
the wavevector $k_z$ satisfying the coupled
Eqs.~(\protect\ref{eq1}),(\protect\ref{eq2})
traces a curved path in the fcc Brillouin zone,
due to the strongly dispersive properties of the photonic crystal.

\bibitem{note_lklu}
Notice that if the reduced DOS was summed over both
positive and negative $k_z$, it would become identical
for the LK and LU directions. However, the two orientations
are physically inequivalent and give rise to different
transmission spectra, thus it is useful to define the
reduced DOS by summing over positive $k_z$
as done in Eq.(\protect\ref{dos}).

\bibitem{whittaker99}
D.M. Whittaker and I.S. Culshaw,
Phys. Rev. B {\bf 60}, 2610 (1999).

\bibitem{note_cylinder}
Since the scattering-matrix method of Ref.~\onlinecite{whittaker99}
applies to a structures consisting of patterned layers
which are homogeneous along a specified ($z$) direction,
the dielectric spheres in the opal structure have been approximated
with cylindrical layers.
The results shown in Fig.~4c,d are obtained by averaging
over calculations with a number of periods in the [111] direction
ranging from 4 to 10, in order to smooth out finite-size oscillations.


\bibitem{dav1} D. Comoretto, R. Grassi, F. Marabelli, and L.C. Andreani,
            Mat. Sci Eng. C {\bf 23}, 61 (2003);
D. Comoretto, F. Marabelli, C. Soci, M. Galli, E. Pavarini, M. Patrini
and L.C. Andreani, Synt. Met. {\bf 139}, 633 (2003).

\bibitem{dav2} D. Comoretto, D. Cavallo, G. Dellepiane, R. Grassi,
F. Marabelli, L.C. Andreani, C.J. Brabec, A. Andreev, and A.A. Zakhidov,
Mat. Res. Soc. Symp. Proc. {\bf 708}, BB10.19.1 (2002).
\end{thebibliography}
\end{document}